\documentclass[twocolumn,showpacs,preprintnumbers,amsmath,amssymb]{revtex4}

\usepackage{graphicx}
\usepackage{dcolumn}
\usepackage{bm}

\pdfoutput=1

\begin{document}


\title{Semiclassical
Hartree-Fock theory
of a  rotating Bose-Einstein condensation}


\author{Ahmed S. Hassan}  
\email{ahmedhassan117@yahoo.com}
\author{Azza M. El-Badry}
\author{Shemi S. M. Soliman}
\affiliation{Department of Physics, Faculty of Science,  Minia  University, El Minia, Egypt.}

\begin{abstract}
In this paper, we  investigate the thermodynamic behavior of a rotating Bose-Einstein condensation with non-zero interatomic interactions theoretically. The analysis relies on a semiclassical Hartree-Fock approximation where an integral is performed over the phase space and function of the grand canonical ensemble is derived. Subsequently, we use this result to derive several thermodynamic quantities including the condensate fraction, critical temperature, entropy and heat capacity.
 Thereby, we investigate the effect of the rotation rate and interactions parameter on the thermodynamic behavior. The role of finite size is discussed. 
Our approach can be extended to consider the rotating condensate in optical potential. 
\end{abstract}



\date{\today}

\maketitle

\section{Introduction}
\label{intr} 
One of the most remarkable characteristics of a Bose-Einstein condensate (BEC) is its response to rotate with  superfluid nature \cite{madison,mattews,codd}.
This property makes the inclusion of interatomic interactions an essential requirement for the correct description of the system. 
In spite of the basic phenomenon of BEC may be illustrated with an ideal gas, it became clear that the interatomic interactions play a predominant role in determining both the qualitative and the quantitative properties of the ultracold gases, especially, for temperatures below the transition.

However,
the statistical mechanics of the interacting system  remains unsolvable and one has to resort to approximated schemes such as Hartree-Fock (HF) approximations \cite{dal}. In this respect, the semiclassical HF  description has produced excellent agreement with experimentally measured equilibrium quantities \cite{gio}. While HF is still very useful, since it does take into account interatomic interactions and being a mean-field theory, it should give rise to a correct qualitative picture of the phase transition.  Many open questions remain predominantly related to exploring the effects of interatomic interactions on the behavior of this system under different circumstances \cite{tammuz}. These include:  the effect of interaction on the BEC transition temperature \cite{strn,sss};
the heat capacity for the system, which enabled us to discuss  the order of phase transition \cite{ket1,Shiozaki}  and the entropy of the system \cite{Olf}, which  required  to investigate the adiabatic cooling of the boson system in lattice to magnetic ordering.

\par
 In the present paper, motivated by the careful study of harmonically confined Bose gas in a rotating trap \cite{Kling,fetter,qiang}, we  employed a developed  semiclassical approximation. The sum over the discrete spectrum for the grand thermodynamic potential is converted into  an integral over phase space. Performing the  integral over this phase space required to calculate many system parameters, such as the condensate density, the effective potential as well as the chemical potential. However, all of them may be self-consistently parametrized   using the Hartree-Fock approximation \cite{Sinha,Sandoval,Jae}.
	Using the thermodynamical potential, the condensed fraction, the transition temperature,  entropy  and the heat capacity are calculated.
Our numerical results are calculated by using the trap parameters of Coddington et al. experiment\cite{codd}. 
	The calculated results showed that the thermodynamic properties  depend strongly on the interatomic interaction and the rotation rate on the whole temperature range.

 \par 
The  paper is planned as follows: section two includes  the system definition   and a systematic method for  calculating 
 the accurate thermodynamic potential. 
The thermodynamic quantities are given in section three. Conclusion is given in the last section.
\section{Basic formalism}
We consider a mesoscopic sample of weakly interacting $N$ bosonic atoms of
mass $m$ placed in an axially symmetric harmonic potential,
$
V_{trap}(r_\bot, z) = \frac{m}{2} ( \omega_\bot^2 r_\bot^2 +  \omega_z^2 z^2)
$,
with  $r_\bot^2 = x^2 + y^2$ is the perpendicular radius and $\{\omega_\bot \equiv  \omega_x = \omega_y , \omega_z\}$ are the effective trapping frequencies of the  harmonic  potential. 
The gas
is set in rotation using an anisotropic quadratic potential $V$
in the $xy$ plane, rotating at angular frequency $\Omega$
 around the
$z$ axis. In the rotating frame, this stirring potential reads \cite{bargi,peta}, 
\begin{equation}
V_{rot}(r_\bot, z) = \frac{1}{2} m  [ \omega_\bot^2 (\kappa_+\kappa_-) r_\bot^2 +  \omega_z^2 z^2 ]
\label{eq2}
\end{equation}  
where $ \kappa_\pm = {(1 \mp \alpha)}$ and  $\alpha = \frac{\Omega}{\omega_\bot }$, is the rotation rate.

The Hamiltonian describing the interacting atomic gas in the potential (\ref{eq2})
is given by\cite{cooper}

\begin{equation}
H = \frac{|{\bf { p_\bot}}  - m {\bf \Omega \times r_\bot}|^2}{2m} + \frac{p_z^2}{2m} +  V_{eff}(r_\bot, z),
\label{eq4}
\end{equation}
where $ V_{eff}(r_\bot, z)$ is the effective potential for rotating interacting condensate boson,
\begin{equation}
	V_{eff}(r_\bot, z) = V_{rot}(r_\bot, z)  + 2 g [n_{th} (r_\bot, z) + n_0(r_\bot, z)],  
	\label{eq3-1} 
\end{equation}
with $g = \frac{4\pi\hbar^2 a }{m}$ is the interaction strength, $n_0(r_\bot, z)$ and $n_{th}(r_\bot, z)$ are the density of condensate and thermal atoms in the rotating frame. 


Usually,  BEC is described within the grand canonical ensemble. All relevant thermodynamic quantities can be calculated from partial derivative of the grand potential $q$, which is the logarithm of the grand canonical partition function \cite{kir,pat}.
\begin{equation}
q(\alpha, T) = - \sum_{n=0}^\infty \ln (1 - e^{-\beta (E_n- \mu(\alpha))})
\label{eq-4}
\end{equation}
where $\beta=1/(k_B T)$ and $\mu(\alpha)$ is the chemical potential of the rotating Bose condensate boson. 
 It is convenient to separate out the ground state contribution and expand the logarithm, $\ln(1-y)  = - \sum_{j=1}^\infty \frac{y^j}{j}$, to express $q$ as a sum over
Bose-Einstein distribution\cite{ket1},
$
	N_n = \frac{{\textsc z} e^{-\beta E_n}}{1 - {\textsc z} e^{-\beta E_n}} = \sum_{j=1}^\infty {\textsc z}^j \sum_{n=0}^\infty e^{-j\beta E_n}
	$.
Thus, Eq.(\ref{eq-4}) can be rewritten as,
\begin{eqnarray}
q(\alpha, T) &=&q_{o} + \sum_{j}\frac{{\textsc z}^j}{j} \sum_{n=1}^\infty e^{-j\beta  E_n }  \nonumber\\  
&\equiv&  q_0 + q_{th}
\label{eq4-1}
\end{eqnarray}
where $ q_{o}=-\ln (1-  {\textsc z}) $ is the grand potential for the atoms in the ground state, with  $ {\textsc z} = e^{\beta \mu(\alpha)}$ is the effective fugacity  and $q_{th}$ is the grand potential for thermal atoms.
 
The sum in Eq.(\ref{eq4-1}) cannot be evaluated analytically in a closed form. Another possible way to do
this analysis is to approximate the sum by integral (semiclassical approximation) over the phase space or converting the sum into an integral weighted by an appropriate smooth density of states (DOS), $\rho(E)$. These two approximations required that the condition $K_BT$  is much larger than the energy level spacing of the system.

\section{ Semiclassical approximation}
\subsection{Hartree-Fock  approximation}
The sum over $n$ in Eq.(\ref{eq4-1}) can be converted into an integral over the phase space by replacing the discrete $E_n$ with a continuous
variable $\epsilon({\bf r; p})$ depends on  position $\bf r$ and momentum $\bf p$, which corresponds to the
classical energy associated with the single-particle Hamiltonian for the system given in Eq.(\ref{eq4}). This
does not take into account the contribution from $n = 0$. In three dimensions, there is
on average one quantum state per volume of phase space $(2\pi\hbar)^3$. Integrating over all
phase space and dividing by this factor thus yields  the thermodynamic potential, $q_{th}$, for the atoms that occupied the excited states,
 \cite{Sinha,hau},

\begin{eqnarray}
	q_{th} ({\bf p, r}) &=& -\frac{1}{(2\pi \hbar)^3} \sum_{j=1}^\infty \frac{{\textsc z}^j}{j}  \int {d^2p_\bot dp_z d^2r_\bot dz}\nonumber\\
	&\times&   e^{-j\beta[\frac{|{\bf { p_\bot}}  -  m {\bf  \Omega \times r_\bot}|^2}{2m} + \frac{p_z^2}{2m} +  V_{eff}(r_\bot, z)]}
\label{eq5}
\end{eqnarray}  
After doing the $p$ integration  by making the change of variables 
${\bf p \to p} -  m {\bf \Omega} \times {\bf r}$,
the  integral in Eq.(\ref{eq5}) takes the same form as in the absence of synthetic magnetic field with an effective 
frequencies  $\sqrt{\omega_\bot^2 (\kappa_+\kappa_-) }$ and $\omega_z$, respectively.
 Finally,  the local grand potential is given by
\begin{equation}
q_{th} ({\bf  r}) = \frac{1}{\lambda_{th}^3} \int   \sum_{j=1}^\infty \frac{{\textsc z}^j }{j^{5/2}} \int  e^{- j\beta  V_{eff}(r_\bot, z) } { d^2r_\bot dz} 
	\label{eq5-2} 
\end{equation}
where $\lambda_{th} = \sqrt{\frac{2 \pi \hbar^2}{m k_B T}}$ is the thermal de-Broglie wavelength.
 However, calculating the phase space integral required calculating  some of the system parameters, include the effective potential, chemical potential and the densities of condensate and thermal atoms. The above mentioned parameters can be calculated using the Hartree-Fock approximation.\\

 In the self-consistent Hartree-Fock model, the thermal atoms are treated as a non-interacting
gas with density $n_{th}(r_\bot, z)$ confined by the effective potential $V_{eff}(r_\bot, z)$ given in Eq.(\ref{eq3-1}).
The densities of the thermal  and condensate component are given as a solution of the two coupled equations:		the thermal atoms satisfy Schr\"odinger's equation
\begin{eqnarray}
\Big[ \frac{|{\bf { p_\bot}}  - m  {\bf  \Omega \times r_\bot}|^2}{2m} + \frac{p_z^2}{2m} &+&  V_{eff}(r_\bot, z) \Big] 
 \psi_i(r_\bot, z) \nonumber\\ &=& \epsilon_i \psi_i(r_\bot, z) 
\label{eq7}
\end{eqnarray}
and  the condensate part satisfies the time independent Gross-Pitaevskii equation,
\begin{eqnarray}
\Big[\frac{|{\bf { p_\bot}}  -  m {\bf  \Omega \times r_\bot}|^2}{2m} + \frac{p_z^2}{2m} &+& V_{rot}(r_\bot, z) + g n_0(r_\bot, z) \nonumber\\  + 2 g n_{th} (r_\bot, z)\Big]  \phi(r_\bot, z) &=& \mu(\alpha) \phi(r_\bot, z),
\label{eq8}
\end{eqnarray}
Eq's. (\ref{eq5}), (\ref{eq7}) and (\ref{eq8}) along with the constraint that the total number of atoms $N$ is fixed,
\begin{equation}
	N = \int n_{th}{(r_\bot, z)} d^2r_\bot dz + \int n_0 {(r_\bot, z)} d^2r_\bot dz
\end{equation}
form a closed set of equations which must be solved self-consistently. 


Both the condensate density $n_0 {(r_\bot, z)}$ and  $\mu(\alpha)$ can be calculated from the time independent Gross-Pitaevskii equation for the condensate part, Eq.(\ref{eq8}). Moreover, the situation may be simplified by taking advantage of the small density of the thermal component (at very low temperature, this requirement may be achieved). In this case, the effect of thermal atoms on the condensate can be  neglected and $n_0(r_\bot, z)$ is given by the
Thomas-Fermi approximation (the kinetic energy term is omitted) of Eq.(\ref{eq8}), leaving an algebraic equation for the condensate density,

\begin{equation}
	n_0(r_\bot, z) = \frac{ \mu(\alpha) - V_{rot}(r_\bot, z)}{g} 
	\label{eq10}
\end{equation}
For all $\mu(\alpha) > V_{rot}(r_\bot, z)$ and $n_0(r_\bot, z)$ = 0 elsewhere.
Substituting from Eq.(\ref{eq2}) into Eq.(\ref{eq10}) leads to,
\begin{equation}
	n_0(r_\bot, z) =  \frac{ \mu(\alpha)}{g} \Big[ 1 - \frac{r_\bot^2}{R_\bot^2(\alpha)} - \frac{z^2}{R_z^2(\alpha)}\Big]  
							\label{eq11}
\end{equation}
where
\begin{eqnarray}
	R_\bot(\alpha) = \sqrt{\frac{ 2\mu(\alpha)}{m   \omega_\bot^2  (\kappa_+\kappa_-)}}\ {\rm and}\  
	   R_z(\alpha) = \sqrt{\frac{ 2\mu(\alpha)}{m  \omega_z^2}},
		\label{eq12} 
\end{eqnarray}
is the Thomas-Fermi radius at which the condensate density
drops to zero along the $r_\bot$ or $z$ axis. Both $R_\bot(\alpha)$ and $R_z(\alpha)$ accounted for the condensate radius in terms of the trap parameters. 
These two radius can be expressed in terms of the condensate number of atoms through the relation between  $\mu(\alpha)$  and $N_0$.
 The relation between $\mu(\alpha)$  and $N_0$ may be founded by integrating
(\ref{eq11}) over the ellipsoid with semi-axes $R_\bot$ and $R_z$,

\begin{eqnarray}
	N_0 &=& \int n_0(r_\bot, z) d^2r_\bot dz \nonumber\\
	 	&=&  \frac{8\pi}{15} \frac{\mu(\alpha)}{g} (R_\bot^2(\alpha) R_z(\alpha)) = \frac{8\pi}{15} \frac{\mu(\alpha)}{g} {\bar R}^3(\alpha)
	\label{eq12-1}
\end{eqnarray}
$\bar R(\alpha)$ is representing the geometric mean $(R_\bot^2(\alpha) R_z(\alpha))^{1/3}$.
Eq.(\ref{eq12-1}) can be inverted to give ${\mu(\alpha)}$ in terms of $N_0$ such as
\begin{equation}
	\mu(\alpha)  = \frac{1}{2} \hbar \omega_g \Big( \frac{15 N_0 a}{a_{har}}  \big)^{2/5}  (\kappa_+\kappa_-)^{\frac{2}{5}} =  \mu(0)  (\kappa_+\kappa_-)^{\frac{2}{5}}
	\label{eq13} 
\end{equation}
where
 $\mu(0) = \frac{1}{2} \hbar \omega_g \Big( \frac{15 N_0 a}{a_{har}}  \big)^{2/5}$
 is the chemical potential for non rotating condensate, $a$ is the s-wave scattering length, $a_{har}=\sqrt{\hbar/m \omega_g}$ and $\omega_g  = (\omega_\bot^2 \omega_z)^{1/3}$.

Further, within the same approximation the effective potential is simply given by
\begin{eqnarray}
	V_{eff}(r_\bot, z) &=& V_{rot}(r_\bot, z)  + 2 g  n_0(r_\bot, z), \nonumber\\
	       &=& | V_{rot}(r_\bot, z)  - \mu(\alpha)| + \mu(\alpha) 
				\label{eq14}
\end{eqnarray}
 Eq.(\ref{eq14}) shows that  the condensate density is drastically altered from the ideal case, reflecting that
the shape of the confining potential  has a three-dimensional `Mexican-hat' shape \cite{bretin}. Moreover, $\mu(\alpha)$ is  the relevant energy scale parameterizing the effects of interactions,
up to the point in the trap where $ \mu(\alpha) = V_{rot}(r_\bot, z)$. 


Finally, in order to calculate the integral given in  Eq.(\ref{eq5-2}), we follow  the Hadzibabic and co-worker \cite{tammuz} approach's and consider the same approximation.
This approach consider that   (compared with $\mu(\alpha)/k_B$) the
majority of thermal atoms lie outside the  condensate in the region where $V_{eff}(r_\bot, z) > \mu(\alpha)$
and $ V_{eff}(r_\bot, z) = V_{rot}(r_\bot, z)$, for relatively high temperature.  Therefore, it is reasonable to approximate  the full effective potential
as the bare trapping potential and consider only the region outside the condensate, i. e.

\begin{eqnarray}
 q_{th} ({\bf  r}) &=& \frac{1}{\lambda_{th}^3} \sum_{j=1}^\infty \frac{{\textsc z}^j }{j^{5/2}} \int  e^{-j\beta  V_{rot}(r_\bot, z) } d^2r_\bot dz\nonumber\\
 &=&  \frac{1}{\lambda_{th}^3}\sum_{j=1}^\infty \frac{1 }{j^{5/2}} \int  e^{-j\beta(\frac{1}{2} m  [\omega_\bot^2 (\kappa_+\kappa_-) r_\bot^2 +  \omega_z^2 z^2 ]  - \mu(\alpha))}\nonumber \\ & & \hskip 3cm  d^2r_\bot dz
	\label{eq15} 
\end{eqnarray} 
introducing a thermal radii, equivalent to the Thomas-Fermi radii given in Eq.(\ref{eq12}), which fixed the maximum value of the chemical potential compared to $k_B T$, 
\begin{eqnarray}
R_\bot'(T) &=& \sqrt{\frac{2k_B T}{ m\omega_\bot^2 (\kappa_+\kappa_-)}},\ 
R_z'(T)	= \sqrt{\frac{2k_B T}{ m\omega_z^2}},
\end{eqnarray}
these radii are  equivalent to  the condensate Thomas-Fermi radii at  which the thermal density
drops to zero along $T \to 0$. Overall, the aspect ratio for the thermal density has the same behavior for the condensate density,
\begin{equation}
	\frac{R_z'(T)}{R_\bot'(T)} = (\kappa_+\kappa_-)^{1/2}
\end{equation}

In terms of $R_\bot'$ and $R_z'$, Eq.(\ref{eq15}) becomes,
\begin{eqnarray}
q_{th} ({\bf  r}) &=&   \frac{1}{\lambda_{th}^3}\sum_{j=1}^\infty \frac{1 }{j^{5/2}} \int  e^{-j\big(  \frac{r_\bot^2}{{R_\bot'}^2} +  \frac{{z}^2}{{R_z'}^2}  - \alpha_0\big)}  d^2r_\bot dz \nonumber\\
 &=& 4\pi \frac{{R_\bot'}^2 R_z'}{\lambda_{th}^3}  \sum_{j=1}^\infty \frac{1 }{j^{5/2}} \int_{\sqrt{\alpha_0}}^\infty R^2  e^{-j(R^2 - \alpha_0)}  dR\nonumber\\
	\label{eq15-1}
\end{eqnarray} 
where the factor $4\pi$ is due to the integration over the angles and
 \begin{eqnarray}
 \alpha_0 &=&   \frac{\mu(\alpha)}{k_B T},\ \
	R^2 = \frac{r_\bot^2}{{R_\bot'}^2} + \frac{{z}^2}{{R_z'}^2},
\end{eqnarray} 
 it is sensible to introduce the variable $Q$, where
\begin{equation}
{Q^2} = {R^2} -  \alpha_0
\end{equation}
to rewrite Eq.(\ref{eq15-1}) as
\begin{eqnarray}
q_{th} ({\bf  r})   &=& 4\pi \frac{{R_\bot'}^2 R_z'}{\lambda_{th}^3}  \sum_{j=1}^\infty \frac{1 }{j^{5/2}}  \int_0^\infty Q^2 \Big(1+ \frac{\alpha_0}{Q^2}\Big)^{\frac{1}{2}} \nonumber\\ 
                 &\times& e^{-j \frac{Q^2}{2}}  dQ \nonumber\\ 
&=& 4\pi \frac{{R_\bot'}^2 R_z'}{\lambda_{th}^3} \sum_{j=1}^\infty \frac{1 }{j^{5/2}}   \int_0^\infty  (Q^2 + \frac{\alpha_0}{2})  e^{-j {Q^2}}  dQ \nonumber\\ 
\label{eq106} 
\end{eqnarray}
where the binomial expansion has been evaluated to first order in $\alpha_0$ .
Evaluating the Gaussian integral in Eq.(\ref{eq106}) gives
\begin{eqnarray}
q_{th} ({\bf  r})  &=& 4\pi \frac{{R_\bot'}^2 R_z'}{\lambda_{th}^3}  \sum_{j=1}^\infty \frac{1 }{j^{5/2}} \Big(  \frac{\sqrt{\pi}/4}{j^{3/2}} +  \frac{\sqrt{\pi}/4}{j^{1/2}}\ \alpha_0  \Big) \nonumber\\
&=& (2\pi)^{3/2} \frac{{R_\bot'}^2 R_z'}{\lambda_{th}^3}  \sum_{j=1}^\infty  \Big(  \frac{1 }{j^4} +  \frac{1 }{j^3}\  \alpha_0  \Big) \nonumber\\
&=& \frac{1}{\kappa_+\kappa_-} \Big(\frac{k_B T}{\hbar \omega_g}\Big)^3 \Big( \zeta(4)  + \alpha_0 \zeta(3) \Big)
	\label{eq16}
\end{eqnarray}
Gathering Eq's(\ref{eq16}) and (\ref{eq4-1}) leads to,
\begin{equation}
q  = q_0 + \frac{1}{\kappa_+\kappa_-} \Big\{ \Big(\frac{k_B T}{\hbar \omega_g}\Big)^3 \zeta(4)  +  \frac{\mu(\alpha)}{k_B T} \Big(\frac{k_B T}{\hbar \omega_g}\Big)^3   \zeta(3)  \Big\}
\label{eqhf}
\end{equation}

Using the same procedure, one can  also obtained results for the total number of particles $N$ \cite{Campbell} and the total energy $E$ \cite{dal}. The total number of particles is given by
 \begin{equation}
N = N_0 + \frac{1}{\kappa_+\kappa_-} \Big\{ \Big(\frac{k_B T}{\hbar \omega_g}\Big)^3 \zeta(3)  +  \frac{\mu(\alpha)}{k_B T} \Big(\frac{k_B T}{\hbar \omega_g}\Big)^3   \zeta(2)  \Big\}
\label{con}
\end{equation} 
While in terms of the $q$-potential,  the total energy is given by $E  =    k_B T^2 \ \Big( \frac{\partial q}{\partial T}\Big)_{ \textsl z}$, thus,
\begin{equation} 
 { E} = { E_0} +  \frac{3 k_B T}{\kappa_+\kappa_-} \Big\{ \Big(\frac{k_B T}{\hbar \omega_g}\Big)^3 \zeta(4)  +  \frac{\mu(\alpha)}{k_B T} \Big(\frac{k_B T}{\hbar \omega_g}\Big)^3   \zeta(3)  \Big\}
	\label{eq6-1} 
\end{equation}

The contribution of the second term in Eq's.(\ref{eqhf}), (\ref{con}) and (\ref{eq6-1}) required to calculate the temperature dependence of the chemical potential.
This dependence  is given by \cite{nar}
\begin{eqnarray}
	\frac{\mu(\alpha)}{k_B T} &=& \frac{\mu(0) }{k_B T}\ (\kappa_+\kappa_-)^{\frac{2}{5}}
	\label{eq13-1}
\end{eqnarray}
 where $\mu(0)$ is the chemical potential for non-rotating boson gas.

The above approximation  is valid for large number of condensate atoms $N_0$ and for
strong repulsive interaction. For small number of particles finite size effect should be considered. However,
the  effect of finite particle number has been found via the density of state approximation.


\subsection{Density of states approximation} 
Another possible way to calculate $q_{th}(\alpha, T)$ is to approximate the sum over the discrete spectrum $E_n$ in Eq.(\ref{eq4-1}) into an  integral weighted by an appropriate  density of states (DOS), $\rho(E)$ \cite{ket1,kir,gro},
\begin{equation}
	q_{th}(\alpha, T) = \sum_{j=1}^\infty\frac{ {\textsc z}^j}{j} \int \rho (E)  e^{-j\beta  E_n} dE
	\label{eq6} 
\end{equation}
 However, calculating $\rho(E)$  in Eq.(\ref{eq6})  required to calculate the spectrum of the  single-particle energy for the Hamiltonian (\ref{eq4}),
which is given by \cite{stock},
 \begin{equation}
 E(n_+,n_-,n_z)  = n_+ \hbar \omega_\bot \kappa_- + n_- \hbar \omega_\bot \kappa_+ + n_z\hbar \omega_z + E_0
 \label{eq18}
\end{equation} 
 where
 $E_0 = \frac{1}{2} \hbar (2 \omega_\bot + \omega_z)$ is the ground state energy  and
 $n_+$, $n_-$ and $n_z$   are  non-negative integers. To follow up the method outlined in our previous paper \cite{ahm1,ahm2,kir,gro}, the accurate DOS  for a many particles system is given by, 
 \begin{eqnarray} 
\rho(\epsilon)  &=& \frac{1}{\kappa_+\kappa_-} \Big\{ \frac{1}{2} \frac{\epsilon^2}{(\hbar \omega_g)^3} + \frac{3}{2} \frac{\bar \omega}{ \omega_g}   \frac{\epsilon}{(\hbar \omega_g)^2} \Big\}
 \label{eq19} 
\end{eqnarray} 
where $\omega_g = (\omega_\bot^2 \omega_z )^{1/3}$ and $ \bar \omega=(2\omega_{\perp} + \omega_z)/3 $.


Substituting Eq.(\ref{eq19}) into (\ref{eq6}), we have the thermodynamical potential for the confined ideal Bose gas in a rotating trap, 
\begin{equation}
q_{th}^{DOS} = \frac{1}{\kappa_+\kappa_-} \Big\{ \Big( \frac{k_B T}{\hbar \omega_g} \Big)^3 g_4({\textsl z}) +  \frac{3}{2} \frac{\bar \omega}{ \omega_g}  \Big( \frac{k_B T}{\hbar \omega_g} \Big)^2 g_3({\textsl z}) \Big\}
	 \label{eq27}
\end{equation}
with $g_k({\textsl z}) = \sum_{j =1}^\infty ({\textsl z}^j/j^k)$ is the usual Bose function.  
Gathering Eq's(\ref{eq27}) and (\ref{eq4-1}) leads to,
\begin{equation}
q^{DOS} = q_0 + \frac{1}{\kappa_+\kappa_-} \Big\{ \Big( \frac{k_B T}{\hbar \omega_g} \Big)^3 g_4({\textsl z}) +  \frac{3}{2} \frac{\bar \omega}{ \omega_g}  \Big( \frac{k_B T}{\hbar \omega_g} \Big)^2 g_3({\textsl z}) \Big\}
\label{eqdos}
\end{equation}
However, using $q^{DOS}$ to calculate the thermodynamic parameters of BEC
is basically identical to that found in our previous work \cite{dal,ahm1,ahm2,su,ahm3,ahm0}, and there is no need to
repeat the analysis here.
\subsection{Critical rotation frequency}
One also must bear in mind that our results are based on the interacting  Bose gas model. As the rotation frequency increase from the slow rotation, there exists a dynamically unstable region of rotating velocities, i.e.
 there exist a critical rotation frequency.
However,	
rotation effect leads to a shift in the radial harmonic oscillator frequencies, bur still fulfill the condition $\hbar \omega_\bot (1 \pm \alpha_c)$, with $\alpha_c$ be the critical rotation rate.  The latter provides the criterion stability of the rotating condensate, it does not necessarily indicate the critical frequency for vortex nucleation. The corresponding thermodynamic rotation rate can be estimated using the relation\cite{fet1}, 
$$\alpha_c \approx 1 - \frac{N a}{\sqrt{8\pi} d_z}$$
where $a$ is the scattering length and $d_z = \sqrt{\frac{\hbar}{m \omega_z}}$ is the ground state spatial extension for the  harmonic potential.

\section{Thermodynamic parameters}
\subsection{Condensate fraction and critical temperature}
Gathering Eq's.(\ref{eqhf}) and (\ref{eq13-1}) leads to,
\begin{equation}
	\frac{N_0}{N} = \Big(\frac{N_0}{N}\Big)_{id} -   K_1(\alpha) {\cal T}^2  
		\label{int}
\end{equation}
where 
\begin{equation}
	\Big(\frac{N_0}{N}\Big)_{id} = 1 - \frac{1}{\kappa_+\kappa_-} {\cal T}^3
\end{equation}  
with  ${\cal T} = \frac{T}{T_0}$ is the normalized temperature and
\begin{equation}
	T_0 = \frac{\hbar \omega_g }{k_B} \Big(\frac{N}{\zeta(3)}\Big)^{\frac{1}{3}} 
\end{equation}
is the transition temperature  of a trapped non-rotating
ideal  gas> The parameter $K_1(\alpha)$ is given by,
\begin{equation}
	K_1(\alpha)  =  \eta\ \big(1 - {\cal T}^3\big)^\frac{2}{5}\  \frac{\zeta(2)}{\zeta(3)}\ (\kappa_+\kappa_-)^{-\frac{3}{5}} 
	\label{eq131}
\end{equation}
with $\zeta$ is the Riemann zeta function. The parameter $\eta$ in Eq.(\ref{eq131}),  first introduced by Stringari et al. \cite{dal,strin}, is determined  by the ratio between the chemical potential at $T=0$ value calculated in Thomas-Fermi approximation and the transition temperature for the non-interacting particles in the same trap i.e. 
$\eta = \frac{\mu_0(T=0)}{K_B T_0} $ ( the typical values for $\eta$ for most
experiments ranges from  0.3 to 0.4.)

In Eq.(\ref{int}), the first term provides the condensate fraction in the thermodynamic limit. The second term, which is vanishes for ${\cal T} > 1$, provides a consistent way for treating the interaction effect \cite{hadz1,hadz2,hadz3,hadz4}.

\begin{figure}
\resizebox{0.50\textwidth}{!}{\includegraphics{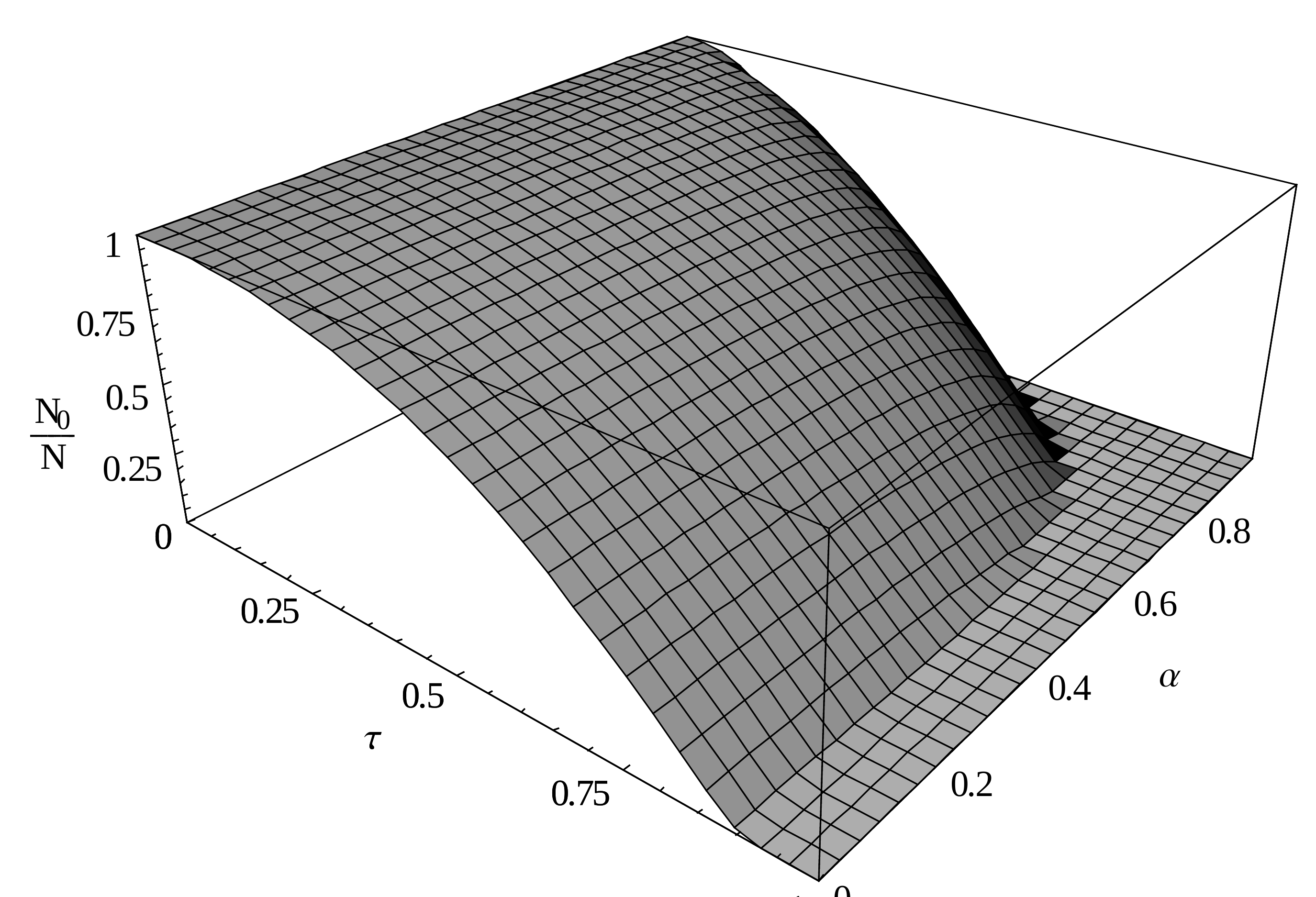}}
\caption{ Condensate fraction versus the  reduced temperature  for different values of  rotation rates $\alpha$ and $\eta = 0.4$. }
\label{f1}
\end{figure}

\begin{figure}
\resizebox{0.50\textwidth}{!}{\includegraphics{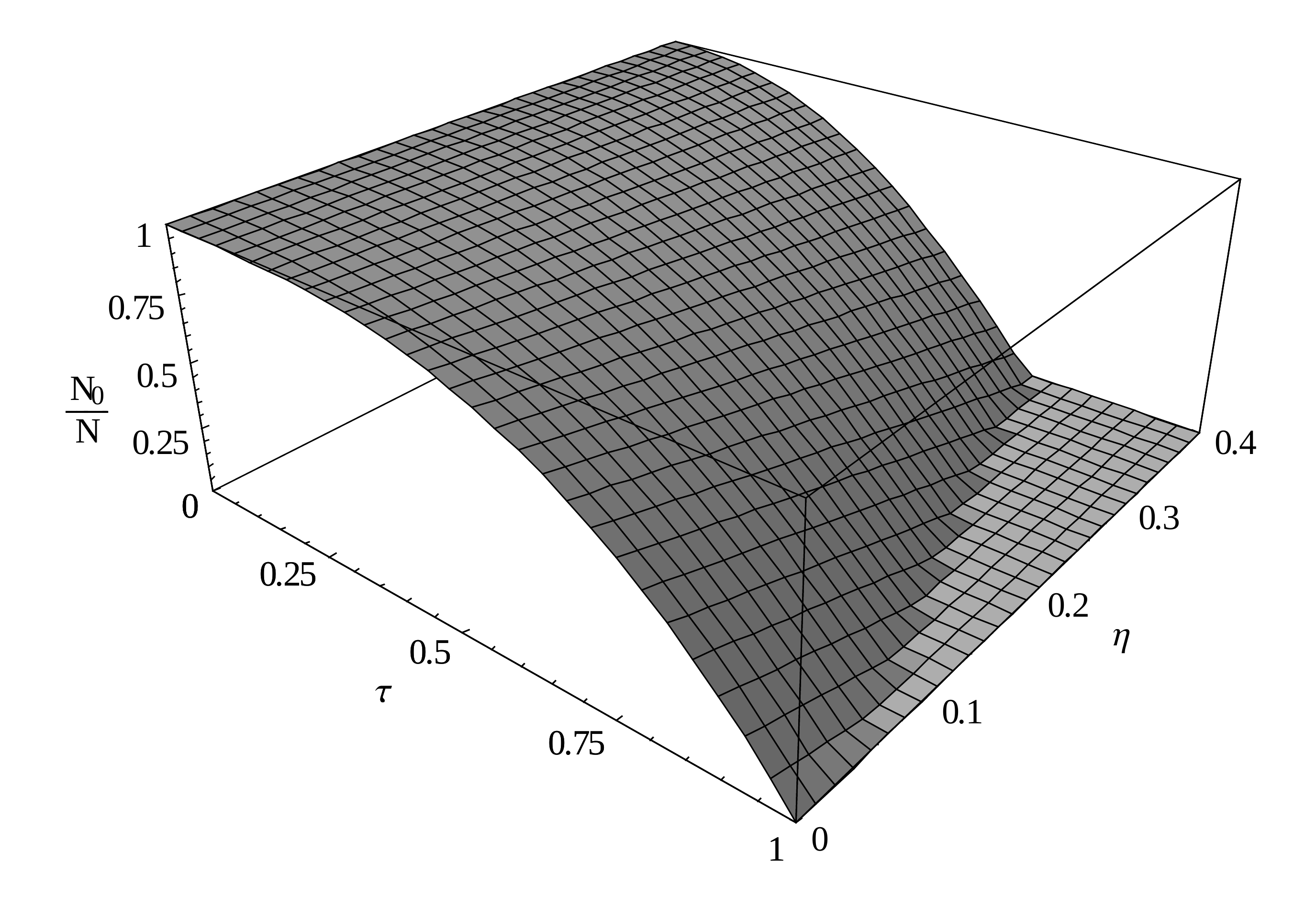}}
\caption{ Condensate fraction versus the  reduced temperature  for different values of  interaction parameter $\eta$ and rotation rate $\alpha = 0.9$. }
\label{f2}
\end{figure}

\begin{figure}
\resizebox{0.50\textwidth}{!}{\includegraphics{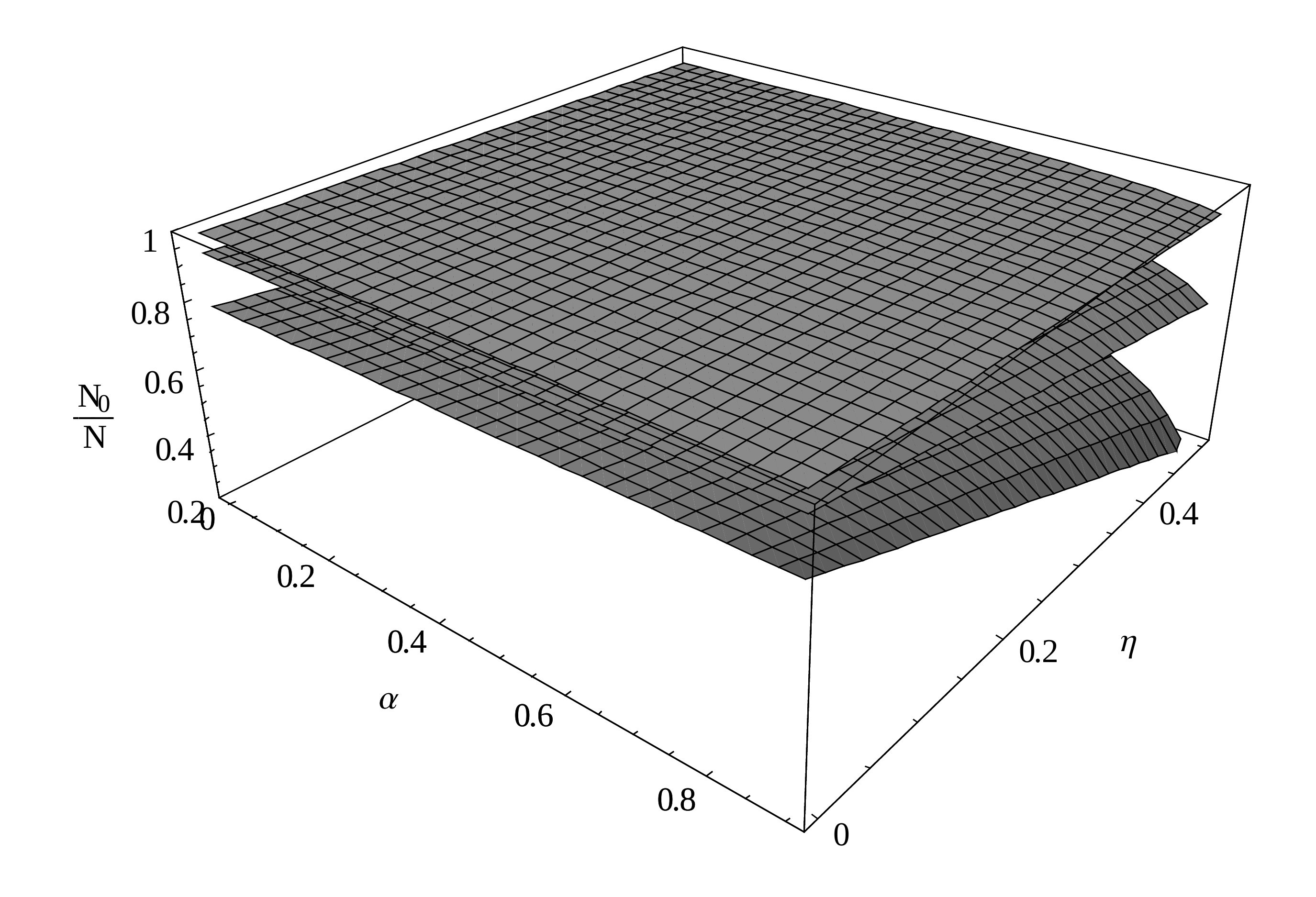}}
\caption{ Condensate fraction versus the  reduced temperature  for different values of  rotation rates $\alpha$ and interaction parameters $\eta$ for ${\cal T} = 0.2, 0.5$ and 0.8. }
\label{f3}
\end{figure}

 In the following, the calculated  results will be considered for  the experimental trap parameters of \cite{codd}:  the  oscillation frequencies are $\omega_x/2\pi = \omega_y/2\pi = 7$ Hz and $\omega_z/2\pi = 13$ Hz.  The interaction parameter for non-rotating gas is taken to be $\eta(0) = 0.4$ and the number of particles is $N = 4.5 \times 10^4$.   In figures \ref{f1}, \ref{f2} and \ref{f3},   the rotation rate and interatomic interaction dependence for the condensate fraction as a function of reduced temperature  are given. 
These figures show that the condensate fraction decreases  as  compared with the non-interacting case due to the
repulsive nature of the interaction. As well as,   for a given values of $N$ and $T$, the values of $N_0$ decreases depending on the rotation rates. 
Which means that,  in a rotating harmonic trap, the condensate gets lost when the rotation frequency comes close to the harmonic trap frequency. 
Thus, the dependence of losing the condensate on the interatomic interaction and the rotation rate $\alpha$ should be taken into consideration for a safe estimate of the critical rotating frequency (rotating frequency required to achieve the vortex state) and critical temperature.

\begin{figure}
\resizebox{0.50\textwidth}{!}{\includegraphics{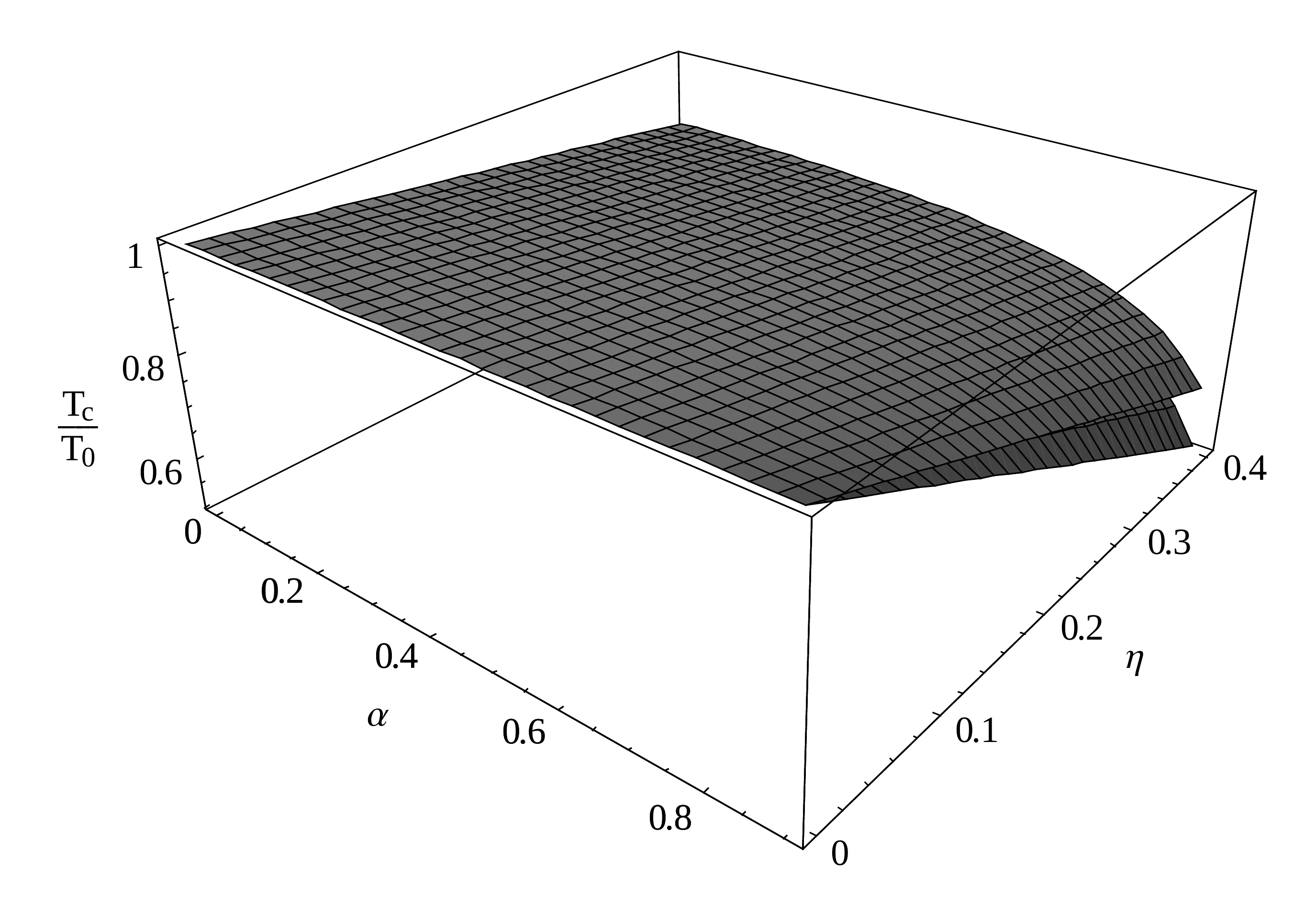}}
\caption{ Critical temperature $T_c$, scaled by the non-rotating transition temperature $T_0$ for an interacting system, as a function of rotation rates $\alpha$. }
\label{f4}
\end{figure}

  The second term in Eq.(\ref{int})   leads to a reduction of the condensate fraction, as well as, it  affected the transition temperature.
   This effect can be seen more clearly by calculating the critical temperature $T_c $.
The latter is obtained as usual \cite{ket1,kir,gro} by setting $N_0/N$ in  Eq.(\ref{int}) equal to zero, thus 
\begin{equation}
T_c  =  T_0 [ 1-   \frac{1}{3} K_1(\alpha) ] 
	\label{eq180}
\end{equation}
In the thermodynamic limit,  the parameter $K_1(\alpha)$ vanishes and the critical temperature reduced to ideal Bose gas in a non-rotating frame $ T_0$.
Eq.(\ref{eq180})  enabled us to investigate  the effects of the  rotation   on $T_c(\Omega)$ in presence of the interatomic  interaction.
 Indeed, in figure \ref{f4}, the normalized critical temperature $(T_c(\Omega)/ T_0)$ is represented graphically as a function of rotation rate $\alpha$ and interaction effect $\eta$.
 This figure shows that the critical temperature $T_c$ decreases  as  compared with the non-interacting case due to the
repulsive nature of the interaction. 

 
\subsection{Entropy of the system}  
A major goal in the field of degenerate quantum gases is to reach a suitable very low temperature. Such low temperatures are necessary to reach phases relevant to condensed matter physics, such as quantum magnetism. However, to ascertain whether a given quantum phases is accessible, it is convenient to focus on its entropy, rather than temperature. Thus, it is important to determine and investigate the entropy-temperature curves\cite{Blakie}. The behavior of these curves is used in analyzing the  process of  adiabatic cooling \cite{Catani,Schachenmayer,Olf}.

 For the rotating condensate,
the normalized entropy per  particle  is given by,
 \begin{equation}
\frac{S}{N k_B} =  \frac{q}{N} + \frac{ E}{Nk_BT} - \frac{\mu(\alpha)}{k_BT}
\label{eq1010}
\end{equation}

Following the usual procedure, the thermodynamic potential for one particle in terms of the normalized temperature ${\cal T}$ is 
\begin{equation} 
  \frac{ q}{N} = \frac{q_0}{N} +  \frac{1}{\kappa_+\kappa_-} \frac{\zeta(4)}{\zeta(3)}  {\cal T}^3 
	+  {\frac{\zeta(3)}{\zeta(2)}} K_1(\alpha) {\cal T}^{2} 
  \label{eq60}  
\end{equation}
While in terms of the $q$-potential  the total energy is given by $E  =    k_B T^2 \ \Big( \frac{\partial q}{\partial T}\Big)_{ \textsl z}$, thus
\begin{equation} 
  \frac{ E}{Nk_BT} = \frac{ E_0}{Nk_BT} + \frac{3}{\kappa_+\kappa_-} \frac{\zeta(4)}{\zeta(3)}  {\cal T}^3 
	+  2 {\frac{\zeta(3)}{\zeta(2)}} K_1(\alpha) {\cal T}^{2} 
	\label{eq61} 
\end{equation}

Substituting from  Eq's.(\ref{eq60})    and (\ref{eq61}) in Eq.(\ref{eq1010}) we have,

\begin{equation} 
	\frac{S}{N k_B} =  \frac{S_0}{N k_B} + \frac{4}{\kappa_+\kappa_-} \frac{\zeta(4)}{\zeta(3)}  {\cal T}^3 
	+  {\frac{\zeta(3)}{\zeta(2)}} K_1(\alpha) \big( 3{\cal T}^{2} -  \frac{\kappa_+\kappa_-}{ {\cal T}} \big)
\label{eq1011}
\end{equation} 
where  $S_0$ is the ground state entropy. 
\begin{figure}[bth]
\resizebox{0.45\textwidth}{!}{\includegraphics{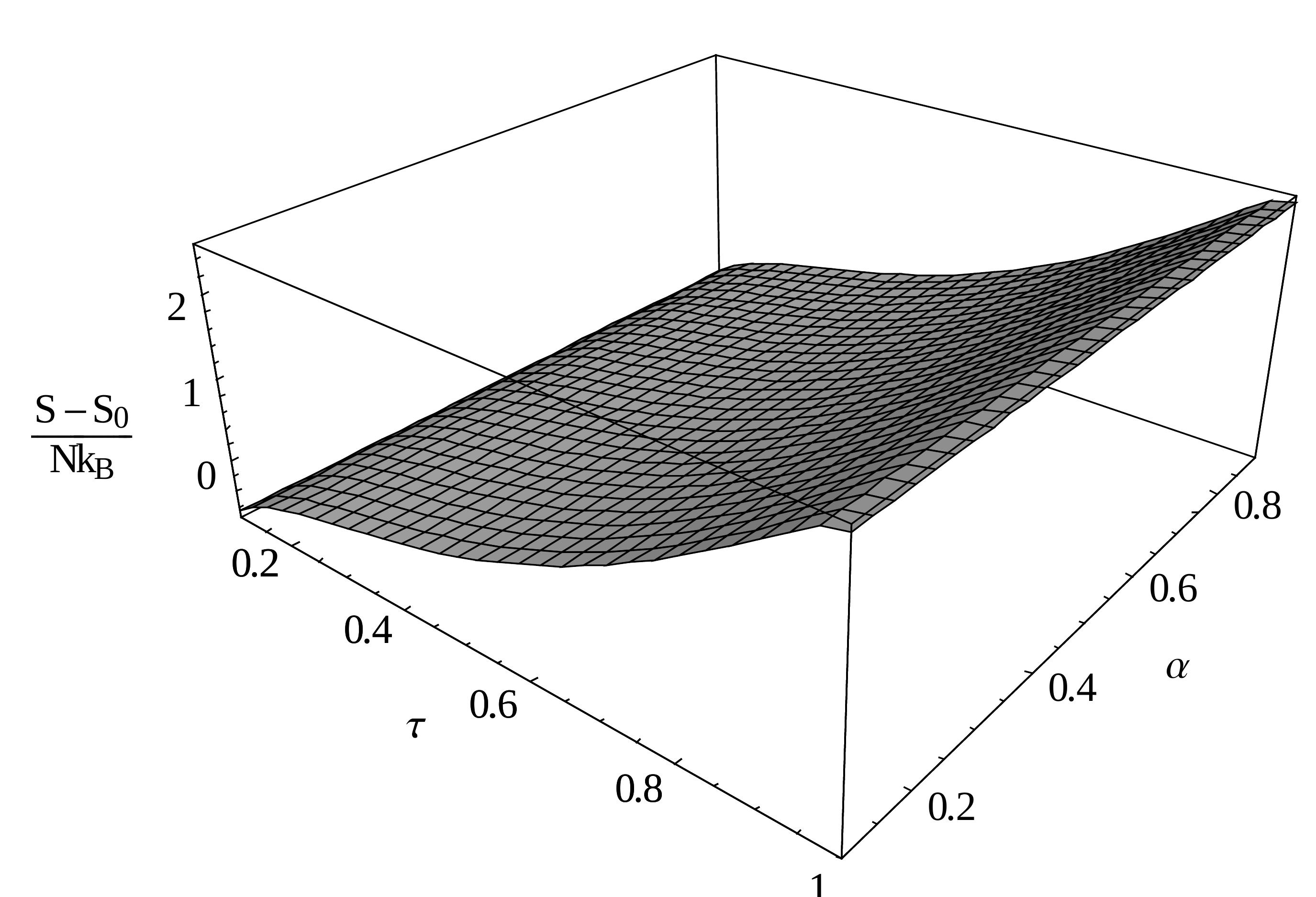}}
\resizebox{0.45\textwidth}{!}{\includegraphics{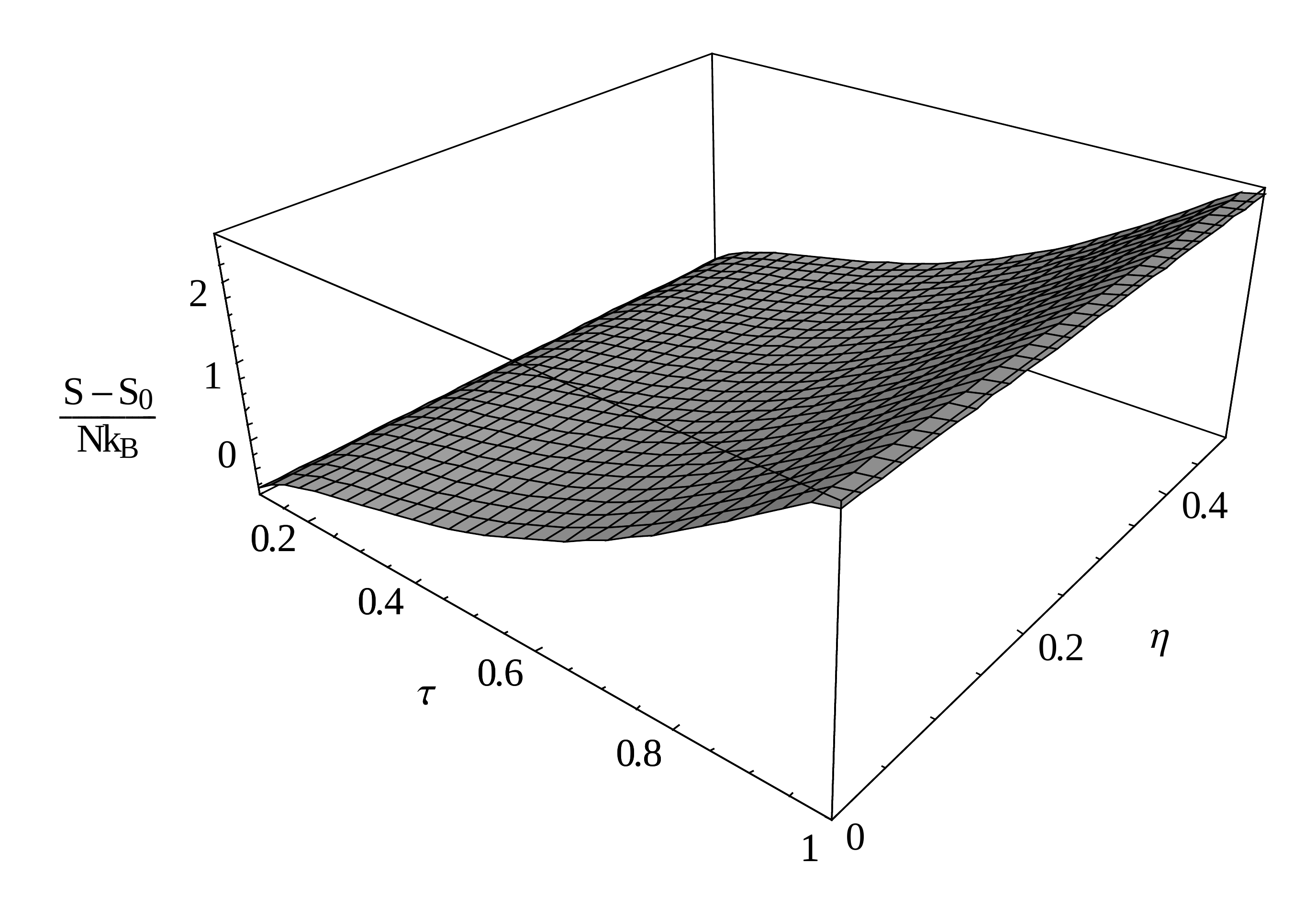}}
\caption{ Entropy versus normalized temperature ${\cal T}$ with $\alpha$ and $\eta$ play as parameters.  
\label{fs}} 
\end{figure} 
In Fig(\ref{fs}),  the entropy versus temperature curves as a function of $\alpha$ and $\eta$ are given. 
These figures show, 
 as it is  expected from standard thermodynamic
arguments, that:   as the temperature  increases  the entropy  has a monotonically  increasing nature everywhere.
Consequently, in order to achieve thermal equilibrium in rotating frame, the trap should contain an asymmetry in the $xy-$plane. Even very small asymmetries are sufficient to ensure thermal equilibrium and safely calculation of the relevant thermodynamic parameters.
However, one of the sensitive quantity to clear up the effects of the rotation and the interatomic interaction  on the condensate is the behavior of the  heat capacity as a function of the reduced temperature.
 

\subsection{Heat capacity}
The essential features of BEC as a phase transition are
clearly exhibited in the behavior of the specific heat, such as in the case of the $\lambda$ point superfluid transition of liquid helium,
which is observed in its heat capacity. 

The heat capacity per a particle at constant volume $\frac{C_V(T)}{Nk_B}$ is of considerable interest. It can be used as an indicator for the order of  the phase transition and for the reduction of the system dimensionality.
\par
  In our approach, 
 \begin{equation} 
 C_V(T) =  \Big( \frac{\partial E}{\partial\textsl T}\Big)_{N, V} =  -    \ \Big( \frac{\partial q}{\partial \beta}\Big)_{ \textsl z},   
  \label{eq24}  
\end{equation}
  However, it is known that  for a given number of atoms, $C_V(T)$ increases to a maximum, then falls rapidly to a saturation value  as $T$ increases greater than $T_0$.
	In such a situation,  we  must take into consideration two different temperature regimes, which are $T$ less or greater than $T_0$.  
	
\par
For ${\cal T} < 1$, the heat capacity is given by 
 \begin{equation}
 \frac{C_{V, {\cal T} < 1}}{N k_B}  = \frac{1}{\kappa_+\kappa_-} \Big\{ 12 {\cal T}^3 \frac{\zeta(4)}{\zeta(3)} +  6\ {K_1(\alpha)} {\cal T}^2 \Big\}
\label{eq35}
\end{equation}
 While the  heat capacity above ${\cal T} > 1$ is given by
 \begin{eqnarray}
  \frac{C_{V, {\cal T} > 1}}{Nk_B }&=& \frac{1}{\kappa_+\kappa_-} \Big\{  12 \frac{\zeta(4)}{\zeta(3)}{\cal T}^3 + 6\  K_1(\alpha) {\cal T}^2 - \Big[3{\cal T}^3 + \nonumber \\
  &2&  
    \frac{\zeta(2)}{\zeta(3)} K_1(\alpha)  {\cal T}^2 \Big] 
   \frac{3 \zeta(3) + 2 K_1(\alpha)  \zeta(2)/{\cal T}}{g_2(z) + K_1(\alpha)   g_1(z)/{\cal T}} \Big\}
  \label{eq37}
\end{eqnarray}

For non-rotating condensation, i.e. $\alpha = 0$, the results previously obtained by  Grossmann and Holthaus \cite{gro} are recovered. While in the thermodynamic limit $N \to \infty (K_1 = 0)$, Eq's.(\ref{eq35}) and (\ref{eq37}) are considerably simplify to,
 \begin{equation}
  \frac{C_{V,T <T_0}^{(\infty)}}{Nk_B } = \frac{12}{\kappa_+\kappa_-} \   \frac{\zeta(4)}{\zeta(3)}{\cal T}^3 
   \label{eq160}   
  \end{equation}
   \begin{equation}
  \frac{C_{V,T >T_0}^{(\infty)}}{Nk_B } = \frac{3}{\kappa_+\kappa_-}\ \Big[ 4\ \frac{\zeta(4)}{g_3(z)} - 3\  \frac{\zeta(3)}{g_2(z)} \Big]
  \label{eq38}
\end{equation}  
Thus, at $T = T_0$ the heat capacity is discontinuous. 
The investigation of the heat capacity jump of
a trapped gas near $T_0$ is important to understand the overall behavior of such phase transition; especially,
for the non-homogeneous confinement case.
However, the magnitude of the jump increases with the rotation rate according to $\frac{6.577}{\kappa_+\kappa_-}$.
This discontinuity  characterizes the phase transition to be of second order according
to the Ehrenfest definition. This means that the system can be described by any potential of our choice. The choice then depends upon the thermodynamic variables you need, rather than upon the transition order. For completeness, in the case of the first order transition, the situation is basically the same. So, we can choose the potential whose variables are more suitable for us. The important difference only arises in the case when a limited portion of the system transforms into a new phase, while the rest of the body stays in the old one. Since we have simultaneously the jump of the solid volume and of the number of particles under the first order transition. So, we cannot fix the volume and the number of particles simultaneously. In this case, it is illegal to use the free energy or other potential whose variables are temperature, volume and the number of particles. We need, instead, to use the so-called thermodynamic-potential with the variables temperature, number of particles and the chemical potential.

Finally, one observes that the heat capacity  Eq.(\ref{eq160})  obeys
the third law of thermodynamics  which demands a vanishing of the heat capacity at zero temperature, and
above $T_0$ is quite linear, in very good agreement with the standard theoretical result: $3k_B/N$ (corresponds
 to the Dulong-Petit law in the very high temperature limit). This interesting general shape of the heat capacity is accepted in the literature\cite{gro,sss,Kling,Jae}.

\par
\begin{figure}[bth]
\resizebox{0.50\textwidth}{!}{\includegraphics{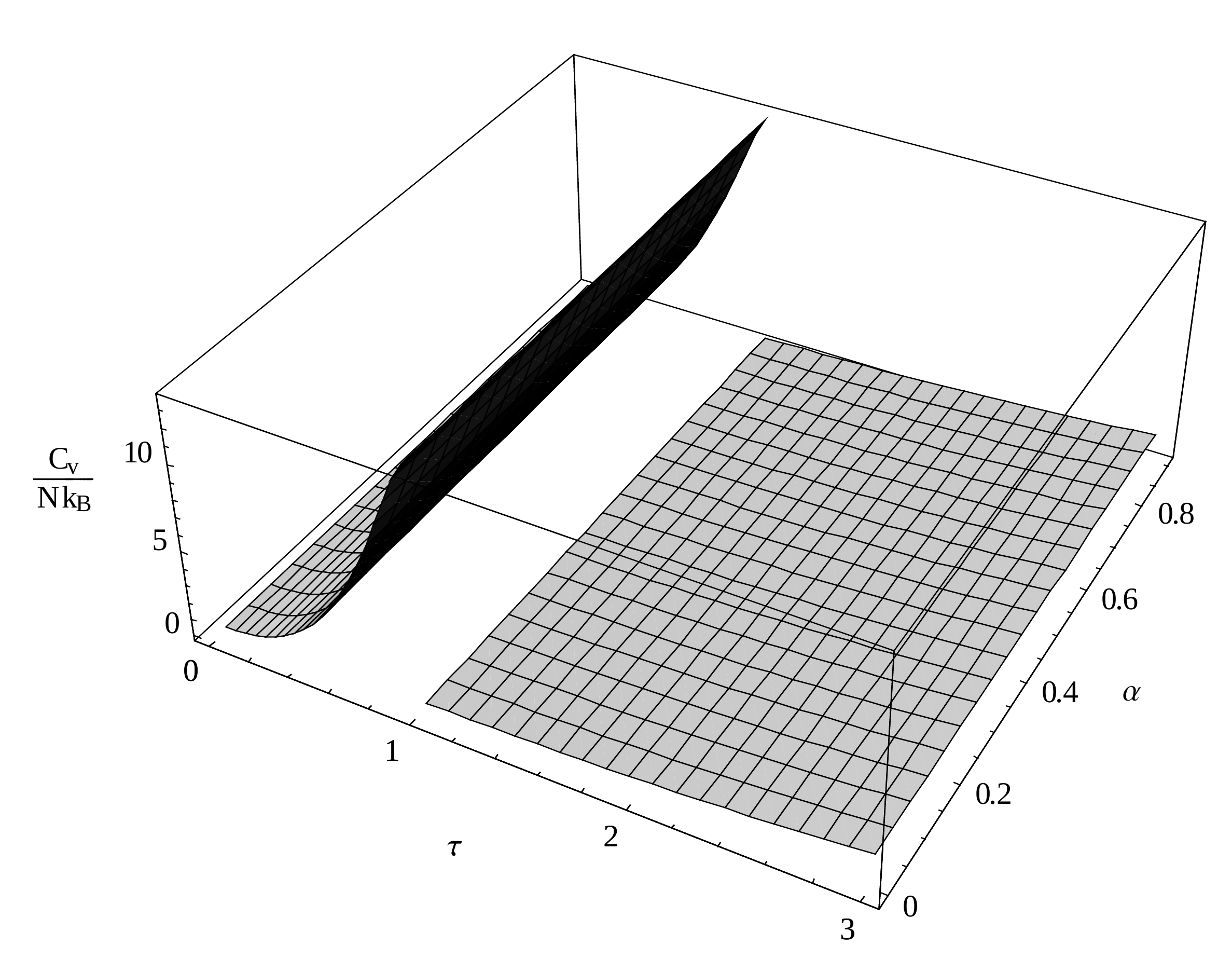}}
\caption{Variation of the  heat capacity $C_V/Nk_B$ with the reduced temperature $T/T_0$ with $\alpha$ play as a parameter. The horizontal solid  line corresponds to the
Dulong-Petit law. }
\label{f7}       
\end{figure}
\begin{figure}[bth]
\resizebox{0.50\textwidth}{!}{\includegraphics{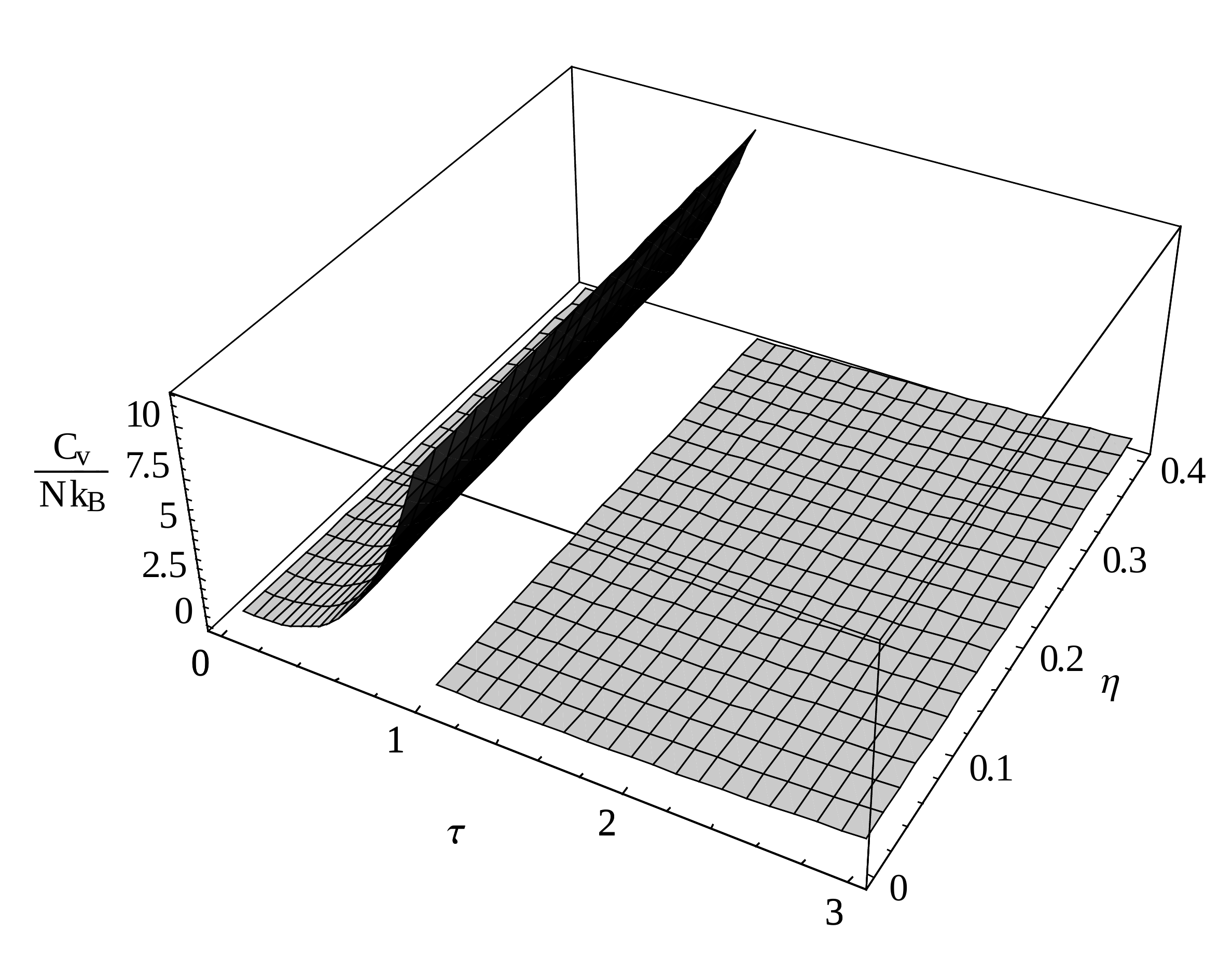}}
\caption{Variation of the  heat capacity $C_V/Nk_B$ with the reduced temperature ${\cal T}$ with $\alpha$ play as a parameter, for $\eta = 0.4.$ }
\label{f8}       
\end{figure}

The results calculated from Eq's.(\ref{eq35}) and (\ref{eq37}) are represented in Fig.(\ref{f7}) and (\ref{f8}) for different  values of $\alpha$ and $\eta$ respectively. The approximation used in \cite{pel} is considered here to calculate Bose function $g_l(z)$ in Eq.(\ref{eq37}).
In Fig.\ref{f7} and Fig.\ref{f8}, we plot the normalized heat capacity $C_V/Nk_B$ versus the normalized temperature, ${\cal T}$ with the rotation rate $\alpha$ and the interaction parameter $\eta$ plays as a parameter.
The heat capacity evolves, starting from zero, with increasing values proportional to the third power of the normalized
temperature, that is: $C_V \propto {\cal T}^3$. At ${\cal T} = 1$ a steep jump takes place
while it goes from ${\cal T} < 1$ to ${\cal T} >1$. Right above the critical temperature, a slow decrease with the temperature is observed
in $C_V$. And, at high temperatures, the heat capacity approaches the temperature independent behavior expected for
the non-interacting Bose gas: $3Nk_B$.

It is interesting to note that signatures of a 
phase transition appear in the   specific heat behavior as a function of ${\cal T}, \alpha$ and $\eta$.
As  {\cal T}  decreases,
the  phase transition, observed at ${\cal T} =1$,
reveals the transition from noncondensed state  to those which is in
condensed phase.


\section{Conclusion} 
In conclusion, by employing the semiclassical Hartree-Fock approximation, we obtain
the analytical expression of the thermodynamic potential
of a  rotating interacting Bose gas in an anisotropic harmonic trap.
Then, the expressions for the condensate
fraction: transition temperature,  entropy and the specific heat
are derived.
The calculated results showed that these thermodynamic quantities
depend on the rotation rate as well as the interatomic interaction for all temperature range.
The critical temperature and
the condensate fraction are decreasing compared with the ideal Bose gas case.
Using $C_v$ as the indicator, we also investigated the  phase transition from the gas phase to condensed phase.
 The method we have outlined here can be extended to study and investigate the thermodynamic properties of a rotating boson gas  in the presence of a combined harmonic lattice potential.

\end{document}